\documentclass[conference]{IEEEtran}
\IEEEoverridecommandlockouts
\PassOptionsToPackage{table,dvipsnames,xcdraw}{xcolor}

\usepackage[utf8]{inputenc} %
\usepackage[T1]{fontenc}    %
\usepackage{hyperref}       %
\usepackage{url}            %
\usepackage{booktabs}       %
\usepackage{nicefrac}       %
\usepackage{microtype}      %

\usepackage{bbm}
\usepackage{dsfont}

\usepackage{cite}
\usepackage{amsmath,amssymb,amsfonts}
\usepackage{algorithmic}
\usepackage{graphicx}
\usepackage{textcomp}

\usepackage[dvipsnames]{xcolor}
\def\BibTeX{{\rm B\kern-.05em{\sc i\kern-.025em b}\kern-.08em
    T\kern-.1667em\lower.7ex\hbox{E}\kern-.125emX}}

\usepackage{adjustbox}
\usepackage{todonotes}
\usepackage{float}
\usepackage{pgfplots}
\pgfplotsset{compat=1.18}
\newif\ifblind
\blindfalse
\newcommand{\blind}[2]{\ifblind #1\else #2\fi}
\newcommand{\CompanyName}{\blind{DOUBLE\_BLIND}{CrowdStrike}}
\newcommand{\PlatformName}{\blind{PLATFORM}{Falcon}}

\makeatletter
\newcommand{\linebreakand}{%
  \end{@IEEEauthorhalign}
  \hfill\mbox{}\par
  \mbox{}\hfill\begin{@IEEEauthorhalign}
}
\makeatother

\begin{document}

\title{Identifying Security Platform Product Abuse with Machine Learning}

\author{
\IEEEauthorblockN{
Shaefer Drew\IEEEauthorrefmark{1},
Michael Brautbar\IEEEauthorrefmark{1},
Paul Knight\IEEEauthorrefmark{2},
Edward Raff\IEEEauthorrefmark{1}, \\
Lana Peric-McDermott\IEEEauthorrefmark{3},
Simran Sarin\IEEEauthorrefmark{1},
Nickolas Machado\IEEEauthorrefmark{4},
Hanna Albright\IEEEauthorrefmark{1},
Vitaly Zaytsev\IEEEauthorrefmark{1}
}
\IEEEauthorblockA{
\IEEEauthorrefmark{1}USA \quad
\IEEEauthorrefmark{2}Canada \quad
\IEEEauthorrefmark{3}Ireland \quad
\IEEEauthorrefmark{4}Brazil \\
\textit{CrowdStrike}\\
\{shaefer.drew, michael.brautbar, paul.knight, edward.raff, lana.pericmcdermott,\\
simran.sarin, nickolas.machado, hanna.albright, vitaly.zaytsev\}@crowdstrike.com
}
}

\maketitle

\begin{abstract}
Product abuse is an individually rare, but growing, problem across the SaaS industry. Highly sophisticated threat actors can misuse security platforms within customer environments or conduct bypass experiments on the product itself. Threat actors can leverage living-off-the-land (LOTL) attacks to avoid using cumbersome, frequently detected malware. Remediating this threat requires collecting multiple data modalities across different types of databases, addressing a cold-start problem in the intrinsic rarity of such sophisticated but dangerous events, and designing within the constraints of real-world deployment (e.g., cost, user behavior, performance, etc). To wit, we provide the first study of such a whole-system defense, especially with respect to a deployed and operational capability. Our results show an increase in product abuse coverage by 35\%, a 30\% reduction in monthly alerts, and adaptability to changes in malicious actors' behavior. We review both the constraints we considered in designing the system to meet operational requirements and a retrospective evaluation of the value of explainable features and counterfactual performance on previously identified attacks.

\end{abstract}

\begin{IEEEkeywords}
product abuse, anomaly detection, risk scoring, machine learning, cybersecurity, explainability
\end{IEEEkeywords}

\section{Introduction}
Some threat actors have exploited enterprise security platforms as attack vectors. An example is a threat actor gaining Single Sign-On (SSO) access to a customer account via a phishing campaign, then using it to log in to the Software-as-a-Service (SaaS) user interface (UI) and disable specific security detections. This poses a significant risk to the security platform and its customers, particularly since the platform is designed to protect them. To protect customers and the product, it is important to detect all forms of product abuse.

Most known existing solutions to security product abuse rely on rules-based alerts. Examples could be "Platform Access from a non-standard browser". In our customer base, these rules haven't generated many high-efficacy leads, often producing too many false positives with minimal coverage of real product abuse. This leads to alert fatigue among investigators assigned to these leads~\cite{gelman_that_2023}. These alert rules are often siloed to single data sources and lack statistical rigor. 

As a solution, \CompanyName{} has developed a machine-learning framework to detect product abuse on security platforms and generate leads for product-abuse investigators. This solution uses anomaly detection and adaptive risk scoring, and is fit to multiple data sources to take full advantage of security platform telemetry and provide suspicious abuse events with more context. Furthermore, it adds explainability to the framework to guide investigators and support their investigation. 
We contribute a study of the system's design process, results, and validation, to help bridge the gap between industry needs and data mining and storage systems. Our goal is to help advance the science of security and the evolving needs in data systems to support this kind of work, which they were often not originally designed for. 

The process begins by engineering features from multiple data sources, including Browser Fingerprint (BFP) data, \textit{interesting events} (which we will define), IP context data, firmographic data, and security tickets. These features then pass through an anomaly-detection algorithm to both filter out normal instances and calculate an anomaly score for downstream risk scoring. Afterward, features from each data source are ranked and ingested into a data-source risk-scoring equation to calculate each data source's risk score. These, along with the anomaly score, are inputs to a final risk-scoring equation that calculates the platform events' final risk. This final risk score has its weights tuned adaptively using Bayesian optimization~\cite{scikit-optimize2021,shahriari2016taking}. Next, a threshold is determined, above which abuse leads are sent to the product security investigators. In addition to the final score, four (4) levels of explainability are implemented to point investigators in the right direction. This involves sending data source risk scores and attributing risk factors to predictions, using SHapley Additive exPlanations (SHAP) to explain local feature contributions for anomaly predictions ~\cite{lundberg2017unified}, and highlighting the overlapping interesting events for the user session. 

Finally, the model is evaluated based on known product abuse incident coverage and alert volume. Our machine learning framework can increase product abuse coverage by 35\% while reducing alert volume by 30\%. From these results, we conclude that this framework is a superior solution to baseline rule-based methods, as it detects more product abuse while reducing analyst fatigue.

\section{Related Work}
The tools needed for system administrators and investigators to perform cybersecurity defensive operations are almost always ``dual use'', in that they can be leveraged for both good (defense) and bad (cyber attacks) purposes~\cite{irwin_double-edged_2018,Raff2020a,raff_cybersecurity_2026}. In this work, we describe, in as much detail as possible without revealing sensitive security details, how we developed a system for detecting threat actors attempting to abuse \CompanyName{} products in malicious ways (e.g., searching for vulnerabilities in \CompanyName{} systems, or testing intrusion techniques to confirm operational success). \textbf{All TPs found through these methods have been remediated.} There is minimal academic literature on this topic at all, let alone from real-world industrial deployment. 

There are some different solutions out there for product abuse in adjacent but unrelated fields, such as Google Apigee abuse detection \cite{apigee}, which detects API abuse using machine learning but doesn't address the complexities and data sources involved in typical EDR product abuse. Our detection has a broader scope, covering the whole EDR platform (UI, API, and Real-Time Response) using a wider range of data sources and techniques tailored to EDR-specific abuse patterns that the API-focused Apigee solution cannot adequately handle.

Non-industry work has been published on sub-components of this problem, but does not consider the aforementioned whole-product scope. Living-off-the-land (LOTL) work often focuses on limited and synthetic data due to the intrinsic difficulty of ``acquiring'' the scripts used by attackers ~\cite{barr-smith_survivalism_2021,ning_survey_2023,trizna_robust_2024,stamp_living-off--land_2022}. Another avenue to produce abuse (though not the only method) is ``account takeover''. This occurs when a malicious actor manages to phish or otherwise obtain control of a pre-existing legitimate account. The methods of how that is technically achieved have also been studied in many domains like finance, e-commerce, and more~\cite{haupert_paying_2017,milka_anatomy_2018,doerfler_evaluating_2019,santoso_detecting_2024,kawase_internet_2019,tao_selective_2018,gao_account_2022}. Because defense-in-depth is necessary in modern security systems, our study focuses on the post-acquisition stage of malicious account acquisition, where multiple systems' data must be integrated to produce an effective, viable solution. 

The first approach attempted in real-world cybersecurity is still having experts write rule-based detection systems. While basic rule-based detection systems can identify simple abuse patterns, they fail to detect sophisticated attacks that mimic legitimate administrative actions, and have long been recognized for producing too many false positives~\cite{bace_nist_2001}. 
The rule-based foundations are still widely used even in modern ML systems, be it using ML to help find new rule candidates~\cite{king_trail_2025,gupta_living_2024}, building reports for investigators~\cite{gao_system_2021,gao_query_2019}, or as a component in a larger ML system~\cite{saqib_gage_2024,song_advancing_2025,sui_bridging_2025}.
Our method addresses the need for an intelligent system that can learn normal behavioral patterns across multiple dimensions, identify subtle deviations that indicate abuse without relying on predefined rules or static thresholds, and explain those deviations to guide analyst investigation.

\section{Data}

In order to detect platform product abuse, we looked towards the telemetry of security data available on \CompanyName{}'s \PlatformName{} platform. 
For product abuse detection, 5 different data sources were chosen to capture user signal and event context. Table \ref{tab:data_source_table} describes these input data sources. The process starts with platform interactions. Each platform interaction collects an event and a browser fingerprint for that device. To reduce the event space, interactions are filtered to only those with the potential for product abuse. For example, \textit{dismissing endpoint detections} as an event could enable product abuse, as threat actors could use it to evade defenses. These \textit{interesting events} are further enriched with IP context, such as VPN information, geography, etc. Customer firmographic information is also joined, detailing high-level information about the customer such as employee count. Finally, our internal ticketing system, consisting of threat hunting leads, managed services, and response tickets, are joined in a way to indicate overlap with the event itself for that user. For example, if \texttt{jane@email.com} is currently under investigation with an open ticket and she is seen dismissing endpoint detections, this will increase the suspicious signal. All of these data sources are combined to provide valuable context for the user and their interactions on the security platform. 

\begin{table}
    \centering
    \adjustbox{max width=\columnwidth}{%
\begin{tabular}{@{}cc@{}}
\toprule
Data Source               & Description                                                                                               \\ \midrule
Browser Fingerprint (BFP) & Browser Device Fingerprints and attributes                                                                \\
Interesting Events        & Platform Events that have potential for abuse                                                             \\
IP Context                & IP behaviors and associations                                                                             \\
Firmographic              & Customer Account Data                                                                                     \\
Security Tickets          & \begin{tabular}[c]{@{}c@{}}Overlapping threat hunting, MDR,\\  and incident response tickets\end{tabular} \\ \bottomrule
\end{tabular}
    }
    \caption{Input Data Sources and Features. Each data source is used to calculate features used as input to anomaly detection and risk scoring. Notably, each comes from a different kind of data store and evolution period, providing long-scale macro (e.g., firmographic) and short-scale (e.g., IP context) information.}
    \label{tab:data_source_table}
\end{table}

\subsection{Features}
For each data source, features and risk indicators were engineered to best discover anomalies and malicious product abuse signals. Because this system is live and used to stop real attacks, we do not specify \textit{all} feature detail to avoid enabling attackers to circumvent it.  

For Browser Fingerprint data, this concerned creating customer and user baselines based on historical activity to differentiate normal from unusual behavior. Inspired by prior user authentication research \cite{MLandS, freeman2016you}, historical device data was used to estimate, for each device attribute and user, the probability that the attribute belongs to that device, producing a value between 0 and 1: closer to 0 indicates the attribute is uncommon for that user, closer to 1 indicates it is frequently observed. For example, a user who has only used Chrome will score high on a Chrome event, but low the first time they use Firefox.

\newcommand{\attr}{\mathcal{A}}  %
\newcommand{\entity}{\mathcal{E}} %
\newcommand{\userid}{u} %

Baselines are created for users by calculating the probability of a device attribute belonging to that user $P(\attr = x \mid u) = \frac{N(\attr = x, u) + \alpha}{N(u) + \beta}$, where $N(\attr = x, u)$ is the number of Browser FingerPrint (BFP) events with attribute $\attr = x$ for user $u$, and $N(u)$ is the total number of BFP events for user $u$.

To avoid zero probabilities, $\alpha = \frac{(\# \text{new } \attr \text{ for user } \userid)}{(\# \text{BFP events})}$ smooths based on how often entirely new attributes appear for a user, while $\beta = 1$ adds a constant to bound the probability between 0 and 1 and prevent infinite values.

Beyond BFP, interesting events were another key data source for the product abuse modeling pipeline. Not only does it filter down the important events, but it also adds many valuable features. As discussed before, interesting events represent events with potential for product abuse if used by malicious actors. Occurrences are converted into count vectors as features. 
IP context features mainly consist of categorical features, such as the VPN in use. These features provide a granular interpretation of the IP and its associations collected from the open internet. Features are categorical (VPN name), continuous, and binary. These features help collect signal towards the IP's reputation. 
Firmographic features concern customer account information. These fields create features themselves but are also combined with other data sources to engineer new features. 
Security ticket features involve features indicating whether or not there is an overlapping ticket for that user and event as well as specific attributions, such as the threat actor seen on the ticket. 
Joining all of these sources enabled substantial context and adjacent signaling to be added to the product abuse model. 

\subsection{Labels}

Due to the rarity of product abuse, labels for the historical corpus were sourced using multiple methods. Below outlines the labeling sources used to mark malicious product abuse events:

\begin{enumerate}
    \item \textbf{Existing Rules-Based True Positives} - Using Product Security's alert API, true positive product abuse events that the existing rules caught were labeled as malicious and joined into the dataset. The high and critical alerts were used to form a baseline to measure against. 
    \item \textbf{Threat Hunting Product Abuse Tickets} - There also exist threat hunting tickets specifically for historical product abuse. We can label certain users as malicious based on these tickets.
    \item \textbf{Emulations} - The Product Security team ran a handful of red teaming exercises on internal accounts, conducting numerous types of product abuse. These samples were labeled and joined into the dataset.
    \item \textbf{Manual Labels} - Manual labels came from manual investigations that didn't start as an automated alert, where Product Security investigators identified product abuse against the platform. These instances were used to label more malicious product abuse events in the dataset. 
\end{enumerate}

Even with the labels collected using the methods above, there was still an extreme imbalance of actual product abuse. Therefore, unsupervised learning methods were preferred.

\section{Method}

\begin{figure}[!h]
    \centering
    \includegraphics[width=1\linewidth]{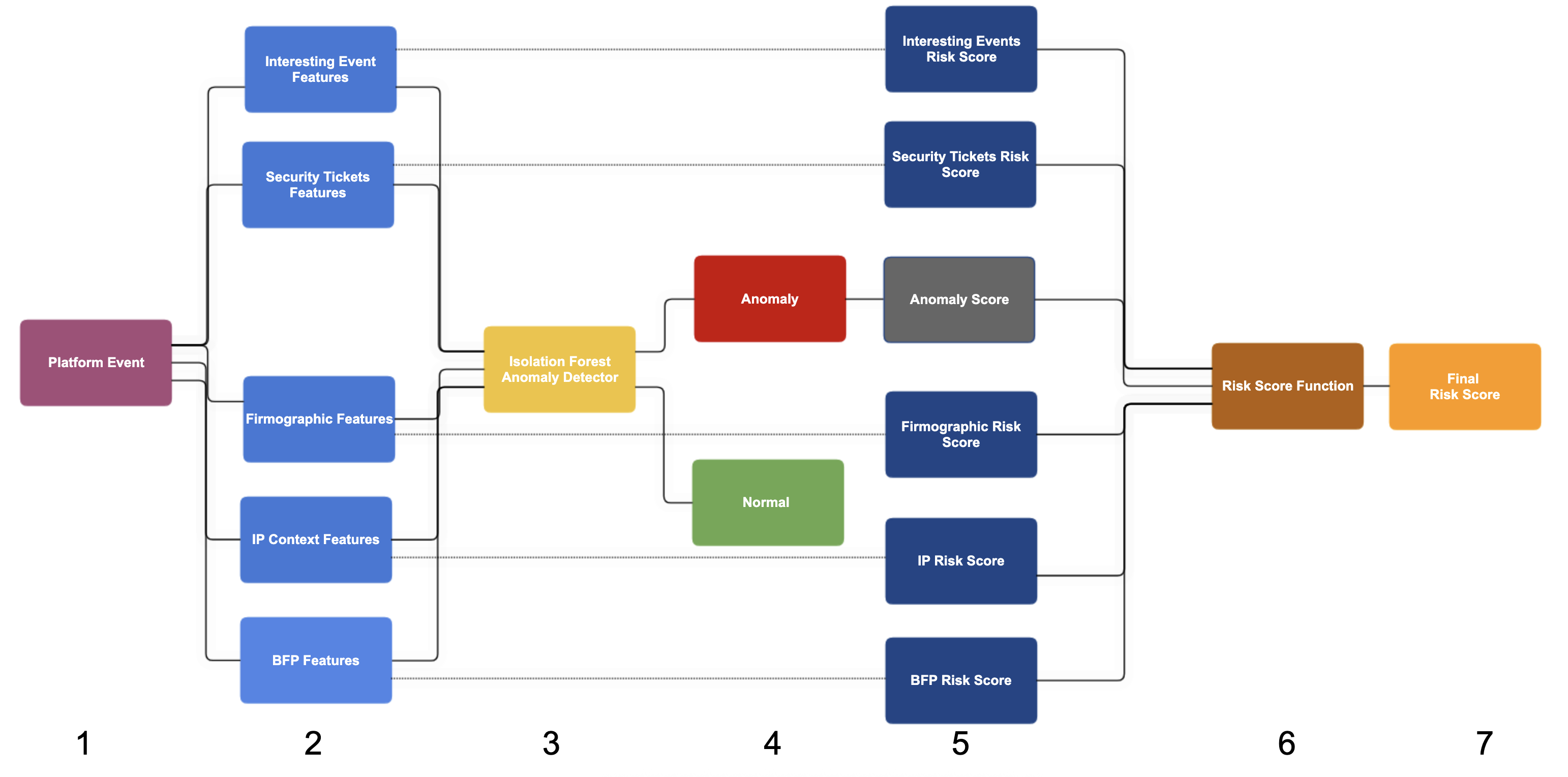}
    \caption{Modeling Diagram. Outlines the ML-based framework of how anomaly detection and risk scoring are combined to generate product abuse leads}
    \label{fig:modeling_diagram}
\end{figure}

\autoref{fig:modeling_diagram} shows the modeling diagram for the product abuse model that generates the final risk score. The process starts with {\color{Purple}platform events for step 1 (\textit{purple})}. Tied to these platform events are Browser Fingerprint data, potentially abusive interesting platform events, IP context data, firmographic data, and correlated security tickets. {\color{RoyalBlue}Features are engineered in step 2 (\textit{blue})} for each of these input data sources. In {\color{Goldenrod}step 3, an anomaly detection model (Isolation Forest ~\cite{liu2008isolation}) (\textit{yellow})} ingests these features and flags the top anomalous events according to a defined contamination ratio threshold. A \textit{contamination ratio} is the proportion of outliers suspected to "contaminate" the data set. For example, a contamination ratio of 0.05 will mark 5\% of the data set as anomalies. In step 4 ({\color{Red} red} and {\color{green} green}), these {\color{Red} filtered proportions of anomalous events (\textit{red})} are sent to step 5, while the {\color{Green} normal events (\textit{green}) are dropped}. In step 5, {\color{MidnightBlue} data source risk scores (\textit{dark blue})} are computed using features from their respective data sources (\textit{dotted lines connecting risk scores to their sources}). The {\color{Gray} anomaly score (\textit{gray})} is also stored in step 5 from the {\color{Goldenrod} anomaly detector (\textit{yellow})}. Both the anomaly score and the data source scores are used as input to the {\color{Brown} step 6 risk score function (\textit{brown})}. The risk score function calculates a weighted risk score from all scores in step 5 and outputs a {\color{Orange} final risk score in step 7 (\textit{orange})}. The highest risk scores above a threshold are sent to product security investigators.

\subsection{Anomaly Detection}
The first modeling portion of the framework involves anomaly detection, which acts as both a pre-filter and an input into the final risk score. We use an Isolation Forest, an unsupervised algorithm that randomly splits trees on features recursively until leaf purity or maximum path length is reached. Anomaly predictions are those with fewer splits, reflecting a statistical difference from the expected number of splits under a random null model \cite{liu2008isolation}.

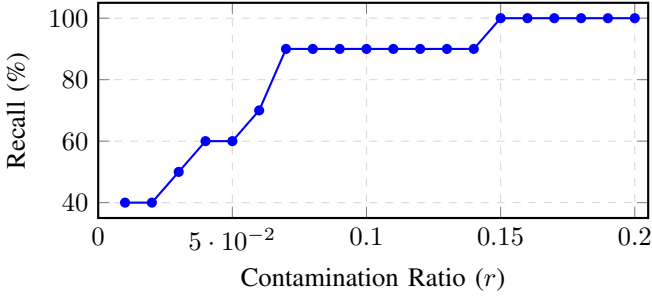
\begin{figure}[!h]
    \centering
    \begin{tikzpicture}
    \begin{axis}[
        width=\columnwidth,
        height=0.5\columnwidth,
        xlabel={Contamination Ratio ($r$)},
        ylabel={Recall (\%)},
        xmin=0, xmax=0.205,
        ymin=35, ymax=105,
        xtick={0,0.05,0.10,0.15,0.20},
        ytick={40,60,80,100},
        grid=major,
        grid style={dashed,gray!30},
        mark size=1.5pt,
        thick,
    ]
    \addplot[blue, mark=*, mark options={solid}] coordinates {
        (0.01,40) (0.02,40) (0.03,50) (0.04,60) (0.05,60)
        (0.06,70) (0.07,90) (0.08,90) (0.09,90) (0.10,90)
        (0.11,90) (0.12,90) (0.13,90) (0.14,90) (0.15,100)
        (0.16,100) (0.17,100) (0.18,100) (0.19,100) (0.20,100)
    };
    \end{axis}
    \end{tikzpicture}
    \caption{Recall vs Contamination Ratio (what fraction of the dataset we choose to mark as anomalous, as ranked by our model). Based on historical data we can acheive 100\% recall considering only 20\% of potential alerts, and still highly effective 40\% at $\leq$ 2.5\%, giving us an effective means to balance analyst availability and ``hunt'' for such rare but dangerous events. }
    \label{fig:contam_ratio_plt}
\end{figure}

For this model, a contamination ratio $r$ was pre-selected based on downstream classification results on a random sample of the data. Recall was plotted in \autoref{fig:contam_ratio_plt} for different values of $r$, showing the trade-off between the proportion of events to label as anomalous vs. the proportion of malicious events captured. All anomaly predictions were filtered through the pipeline; all normal predictions were filtered out. 
The anomaly score for all anomalous instances was stored for further usage. This significantly filtered down the event space and had downstream effects on the final risk score.

\subsection{Adaptive Risk Scoring}
The next part of the pipeline involved adaptive risk scoring. Anomaly detection alone struggles to separate malicious from benign. For example, a rare enterprise VPN may raise the anomaly score despite being a benign signal. Since this problem lacks a large labeled corpus, risk scoring combines \textit{anomalous} signals with \textit{suspicious} signals, beginning with data source risk scores that reflect weighted risk factors. These are then ingested by a final risk function weighting the anomaly score and data-source scores, with scores above a threshold sent as leads for analyst investigation.

\subsubsection{Data Source Risk Scores}
The following equation shows how data-source-level risk scores are calculated using a weighted, tiered dominance approach with exponential decay across risk factor tiers. Since the lack of data makes a pure machine learning approach difficult, we instead work with domain experts to define informed ordinal categories with weights set by hyper-parameter tuning. This proved effective when merged this with the more data-heavy unsupervised approach.

To create the input for the data source risk scores, Product Security investigators ranked the importance of risk features (risk factors) on a scale from Somewhat Good (SG) to Extremely Bad (EB). These ordinal rankings are based on domain expert guidance and reflect their expertise. "Bad" factors increase the data source risk score while "Good" factors decrease the data source risk score. An example could be "Suspicious VPN" falling under "Very Bad". This represents feature categorization via analyst annotation. It is completely separate from the target variable labels.

Table \ref{tab:risk_factors_table} shows how certain features can be ranked as risk factors. Each factor is one-hot boolean encoded with a 0 or a 1 depending on if the factor was hit for a given event.
\begin{table}[!h]
    \centering
    \caption{Risk Factors and Rankings. Examples of risk factor features and their rankings provided by subject matter experts. These are used as input to data source risk scores. Very/Extremely Bad ranks contribute to higher marginal risk increases than slightly bad ranks.}
    \begin{tabular}{|l|l|l|}
        \hline
        \textbf{Risk Factor} & \textbf{Data Source} & \textbf{Rank} \\
        \hline
        Suspicious VPN & IP Context & Very Bad \\
        \hline
        Event 12 (\textit{creating a new user}) & Interesting Events & Slightly Bad\\
        \hline
        Threat Actor Association & Security Tickets & Extremely Bad \\
        \hline
    \end{tabular}
    \label{tab:risk_factors_table}
\end{table}

Given these ranked factors, an equation was needed to combine risk scores in a meaningful way, satisfying specific properties to be viable for deployment and interpretability by maintainers and investigators. Below are the properties used to derive the data source risk equation:

\begin{enumerate}
    \item \textbf{Strict Hierarchy:} Maintains EB \textgreater VB \textgreater B \textgreater SB for bad factors, G \textgreater SG for good factors
    \item \textbf{Dominance Principle:} Highest present tier dominates the calculation, with lower tiers providing marginal contributions.
    \item \textbf{Bounded Output:} Final score always between 0 and 1
    \item \textbf{Monotonically Increasing} with respect to negative factors
    \item \textbf{Monotonically Decreasing} with respect to good factors
    \item \textbf{Concave Growth Curve:} Each factor demonstrates diminishing returns through exponential decay
    \item \textbf{Asymptotic Approach:} Factor contributions approach their maximum limits as counts increase
    \item \textbf{Tunable Sensitivity:} Alpha parameter adjusts how quickly factors reach maximum effect
    \item \textbf{Diminishing Returns:} First instance of each factor has the highest impact, with decreasing marginal impact for additional instances
    \item \textbf{Base Risk Adjustment:} Starts with a configurable baseline that can be tuned to application context
\end{enumerate}

Equation \ref{eq:risk_score} shows how risk scores combine contributions and 
are bounded between 0 and 1 using a clipping function (\textit{property 3}). 
The risk equation takes a tiered dominance approach (\textit{property 2}), 
adding domain-tier and lower-tier contributions and subtracting good 
contributions from the base score. A clipping function is used rather than 
an alternative bounded transformation such as sigmoid to preserve linear relationships between risk 
factors and the final score, maintaining interpretability and the additive 
nature of the contributions. The baseline is configurable; however, it is 
kept constant for this research (\textit{property 10}).
\begin{equation}
R = \max(0, \min(1, b + D + L - G))
\label{eq:risk_score}
\end{equation}

The ordinal factor counts, in decreasing severity, are: \textbf{EB} (Extremely Bad), \textbf{VB} (Very Bad), \textbf{B} (Bad), \textbf{SB} (Somewhat Bad), \textbf{G} (Good), and \textbf{SG} (Somewhat Good). Each tier has a corresponding dominant-tier weight $W_{(\cdot)}$ (e.g., $W_{EB}$) and, where applicable, a lower-tier weight $W_{(\cdot)}^L$ (e.g., $W_{VB}^L$). The remaining parameters are $\alpha$ (sensitivity/decay rate) and $b \geq 0$ (base score).

Equation \ref{eq:d_component} represents the dominant tier contribution and Equation \ref{eq:l_component} the lower-tier marginal contributions. Let the bad tiers be ordered $t_1 > t_2 > t_3 > t_4$ corresponding to (EB, VB, B, SB), and let $k^* = \min\{k : t_k > 0\}$ denote the index of the highest-severity tier present. Each tier's contribution follows an exponential decay $(1 - e^{-\alpha \cdot t_k})$ that is monotonically increasing with diminishing returns (\textit{properties 4,6,8,9}), bounded at 1 as $t_k \to \infty$. The dominant tier receives weight $W_{t_{k^*}}$ while all lower tiers receive marginal weights $W^L_{t_k}$, maintaining strict hierarchy (\textit{property 1}) and dominance (\textit{property 2}). The sensitivity parameter $\alpha$ was kept constant across all datasets.

\begin{equation}
D = W_{t_{k^*}} \cdot (1 - e^{-\alpha \cdot t_{k^*}})
\label{eq:d_component}
\end{equation}

\begin{equation}
L = \sum_{k > k^*} W^L_{t_k} \cdot (1 - e^{-\alpha \cdot t_k})
\label{eq:l_component}
\end{equation}

Equation \ref{eq:g_component} is a weighted equation that combines all the "good" risk factors, in the "Good" tier. It is monotonically increasing with a strict hierarchy of G \textgreater SG (\textit{properties 1, 5}).

\begin{equation}
G = W_{G} \cdot (1 - e^{-\alpha G}) + W_{SG} \cdot (1 - e^{-\alpha SG})
\label{eq:g_component}
\end{equation}

The overall risk score equation \ref{eq:risk_score} combines the Dominant (D), Lower (L), and Good (G) contributions with the base score (b) to create a data source risk score between 0 and 1.

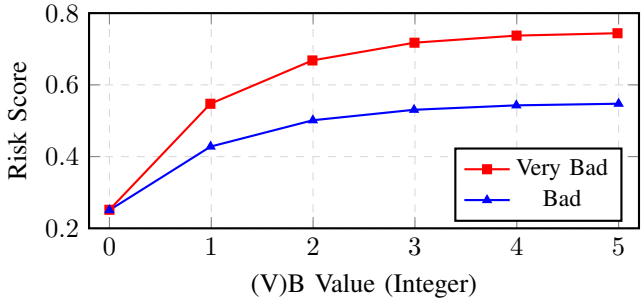
\begin{figure}[!h]
    \centering
    \begin{tikzpicture}
    \begin{axis}[
        width=\columnwidth,
        height=0.5\columnwidth,
        xlabel={(V)B Value (Integer)},
        ylabel={Risk Score},
        xmin=-0.2, xmax=5.2,
        ymin=0.2, ymax=0.8,
        xtick={0,1,2,3,4,5},
        grid=major,
        grid style={dashed,gray!30},
        mark size=1.5pt,
        thick,
        legend style={at={(0.97,0.03)},anchor=south east,font=\small},
    ]
    \addplot[red, mark=square*] table[col sep=comma, header=false] {ImpactofVeryBad.csv};
    \addlegendentry{Very Bad}
    \addplot[blue, mark=triangle*] table[col sep=comma, header=false] {ImpactOfBad.csv};
    \addlegendentry{Bad}
    \end{axis}
    \end{tikzpicture}
    \caption{Visualizes the impact of a risk factor on the initial risk score, based on the number of factors that hit at each rank. Very Bad and Bad are represented in the figure.}
    \label{fig:risk_score_diag1}
\end{figure}

Figure \ref{fig:risk_score_diag1} shows how risk factors are weighted differently and use exponential decay functions. The first factor has the greatest influence on the risk score, with each successive factor contributing a progressively smaller marginal gain or loss. See Appendix A and figure \ref{fig:risk_score_diag4} for the complete charts, which follow the same trend. 

\subsubsection{Final Risk Score}
Once the data source risk scores and the anomaly risk score are calculated and stored, these scores are combined into the final risk score. Intuitively, this risk score is supposed to represent the ratio of malicious to normal. Using a conceptual application inspired by Likelihood Ratio Testing, this is shown by combining the risk scores to create a sense of maliciousness in the numerator while having $1 -$ anomaly score in the denominator representing "normal" \cite{likelihood-ratio_neil}. This creates an amplification effect as the anomaly score approaches 1. The following equation shows how the final product abuse lead risk score is computed by combining all the different data source risk scores and the anomaly score.

Equation \ref{eq:initial_score} shows the weighted combination of all the data source risk scores and anomaly score.
\begin{multline}
\text{scores}_i = W_1 \cdot \text{ARS}_i + W_2 \cdot \text{IPRS}_i + W_3 \cdot \text{STRS}_i \\+ W_4 \cdot \text{IERS}_i + W_5 \cdot \text{BFPRS}_i + W_6 \cdot \text{FRS}_i
\label{eq:initial_score}
\end{multline}
where each feature is continuous  $\in [0,1]$:

\begin{tabular}{ll}
$\text{ARS}_i$ &  : Anomaly Risk Score for alert $i$ \\
$\text{IPRS}_i$ &  : IP Risk Score for alert $i$ \\
$\text{STRS}_i$ &  : Security Ticket Risk Score for alert $i$ \\
$\text{IERS}_i$ & : Interesting Events Risk Score for alert $i$ \\
$\text{BFPRS}_i$ &  : BFP Risk Score for alert $i$ \\
$\text{FRS}_i$ &  : Firmographic Risk Score for alert $i$
\end{tabular}
Equation \ref{eq:normalized_score} then normalizes this score to be between 0 and 1.

\begin{equation}
N_i = \frac{\text{score}_i - \min_j \text{score}_j}{\max_z \text{score}_z- \min_j \text{score}_j}
\label{eq:normalized_score}
\end{equation}

$R_i = \frac{N_i}{1 + W_0(1 - \text{ARS}_i)}$
is how the final risk score is calculated, dividing the normalized numerator by the weighted ($1 - $anomaly score), adding 1 to smooth the score and constrain it to between 0 and 1.

These risk weights are tuned with Bayesian Optimization using Gaussian processes \cite{scikit-optimize2021,shahriari2016taking}. A k-fold bootstrapped approach is taken where a subset of data is randomly sampled, predicted on using a set of weights, and average recall is recorded across each iteration when constrained to a pre-defined flag rate. Bayesian weight tuning allows the risk score to be dynamic, in that we can re-tune the weights with an updated corpus, assigning more weight to data source risk scores that indicate a stronger signal based on the labels. Since the historical dataset has very few malicious labels, the weight-tuning dataset is drawn from the same dataset as the evaluations. This decision was made because of the lack of labeled data, which required a bootstrapped approach, and because of the unsupervised nature of this method. While this may present leakage concerns, the model was still evaluated against live production data that wasn't seen in any of the weight tuning process. Even against the unseen data, the model significantly outperformed baselines. 

Once the final risk scores are calculated, a threshold is created to select only the top leads to show investigators. This threshold is calculated to satisfy volume constraints, setting the threshold to flag only the top leads based on the sample set. This threshold ensures that fewer average monthly leads are flagged vs existing methods on the historical sample.

\subsection{Explainability}
The final risk score alone doesn't meet all the requirements of a lead generator. As such, it was required to be user-friendly for the security investigators. Due to the high dimensionality of input data, simply sending a risk score by itself is insufficient. There is a "cold start" problem that arises when multiple data sources produce these leads, and investigators have no clue where to begin their investigation. Should they start with the browser fingerprint, the event itself, or the IP? A lead generation model needs to tell investigators not just "what" to look at, but "why" and "where" to look first. To speed up investigation time and solve for this cold start problem, explainability was built in as a major component of this model.

\subsubsection{Data Source Risk Attribution}

Starting with the data source and anomaly risk scores, investigators can identify which sources had worse risk factors, giving them a starting point for investigation.

\autoref{tab:risk_scores_tab} shows example risk lead predictions. Product security investigators see the final score and they can also see the data source risk scores that went into the calculation, giving them an idea of where to begin investigating. 
We can also use this risk score setup to explain why events were flagged as malicious. We do this by simply sending a list of which risk factors hit for each event and their corresponding values.

\begin{table}[!h]
\label{tab:risk_scores_tab}
\centering
\caption{Risk Score Analysis by Event. Shows example data source risk scores, the anomaly score, and final risk score for 2 events, allowing investigators to understand which data sources contributed the most to risk.}
\small
\adjustbox{max width=\columnwidth}{%
\begin{tabular}{@{}lcccccc@{}}
\toprule
\textbf{Event} & \textbf{Anomaly} & \textbf{IP} & \textbf{Security} & \textbf{Events} & \textbf{Firmographic} & \textbf{Total} \\
\midrule
a & 0.25 & 0.45 & 0.39 & 0.40 & 0.20 & 0.86 \\
b & 0.30 & 0.78 & 0.44 & 0.34 & 0.60 & 0.89 \\
\bottomrule
\end{tabular}
}
\end{table}

Table \ref{tab:risk_factors_exp} provides an example of explaining leads by highlighting the risk factors and their values that were hit for a single prediction. This user performed interesting event 15 exactly 3 times, has a threat actor associated with an overlapping security ticket, and was using a suspicious browser. In the actual queue, interesting events show the description, not just the mapping. \textit{Interesting event 15} may represent something like a defense evasion or privilege escalation indicator. 

\begin{table}[!h]
    \centering
    \caption{Risk Factors Explanation Example. This demonstrates the risk explanation for a single example product abuse lead, where the factors contributing to the risk score included the user performing interesting event 15 three times, having an associated threat actor in an overlapping security ticket, and using a suspicious browser to perform these actions. These explanations help point investigators in the right direction.}
    \begin{tabular}{|l|c|}
        \hline
        \textbf{Risk Factor} & \textbf{Hit Count} \\
        \hline
        Interesting Event 15 & 3 \\
        \hline
        Threat Actor Association & 1 \\
        \hline
        Suspicious Browser & 1 \\
        \hline
    \end{tabular}
    \label{tab:risk_factors_exp}
\end{table}

\subsubsection{SHAP Anomaly Risk Attribution}

Next, using Isolation Forest's compatibility with SHAP, we flag the top 10 local feature contributions based on Shapley values, a game-theoretic approach connecting optimal credit allocation to local explanations \cite{lundberg2017unified}. SHAP scores for an Isolation Forest analyze marginal feature contributions to path length, with shorter paths yielding higher SHAP values, helping investigators understand why leads were flagged as anomalous.

\begin{figure}[!h]
    \centering
    \includegraphics[width=\columnwidth]{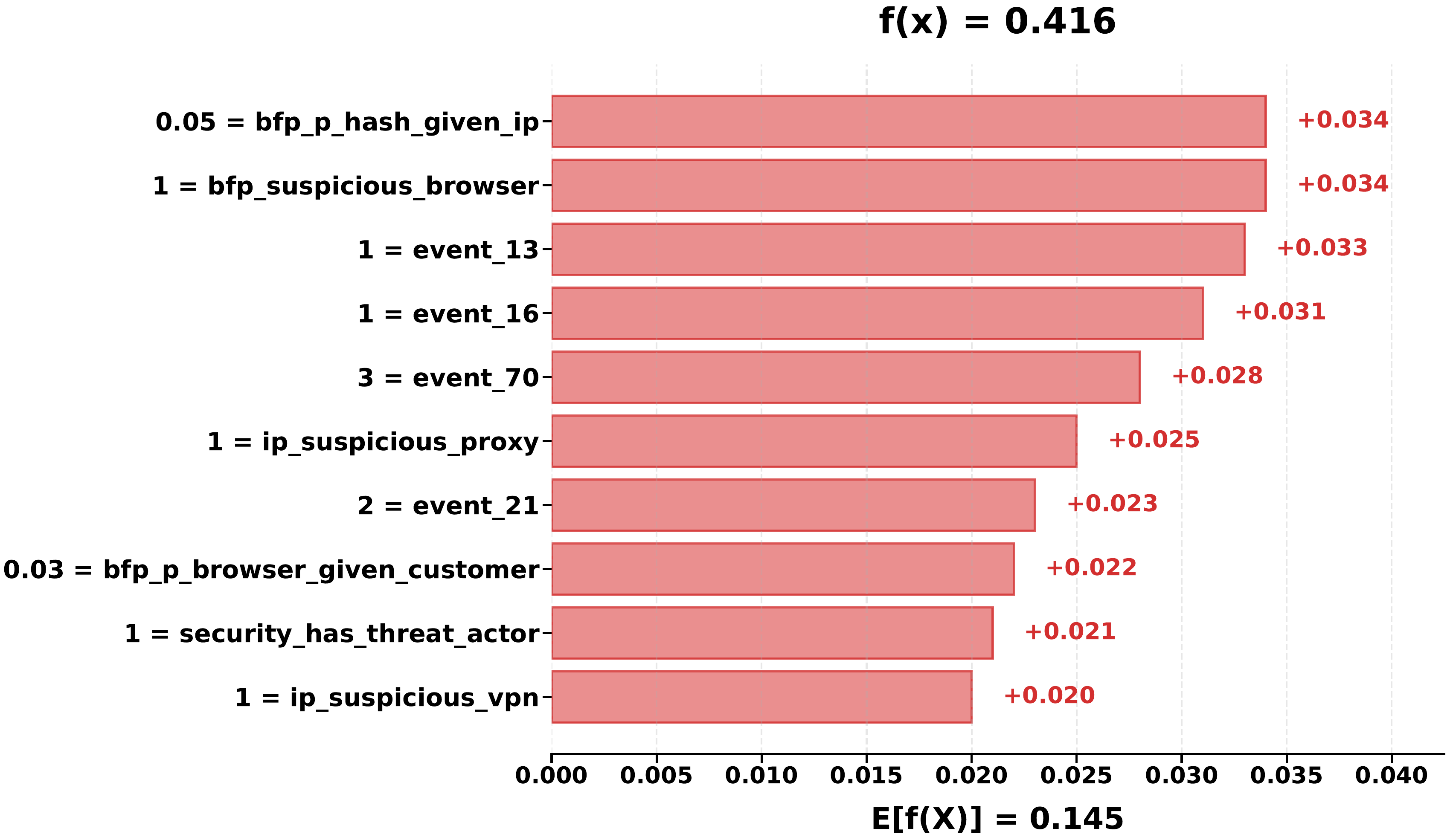}
    \caption{Local Explanation of Top 10 SHAP Features for Isolation Forest, signifying the top feature contributions to the anomaly score. These were shared with investigators (who could resolve events to specific types via a database) who used them to identify discrepancies between initial modeling and how attacks tend to work, and successfully iterate to our current solution. }
    \label{fig:shap_chart}
\end{figure}

Figure \ref{fig:shap_chart} shows the local SHAP explanations for the Isolation Forest model for a single anomaly prediction. This visual uses anomaly score instead of path length. This helps investigators understand why this event may have been flagged as an anomaly. In the example local explanation, among other indicators, it showed the BFP hash had a low probability of belonging to that IP, it was using a suspicious browser, and performed interesting platform event 13. 

\subsubsection{Interesting Events Explainability}

Lastly, interesting events themselves can be flagged as an explanation. Since that was a pre-filter for flagging leads in the first place, investigators are always going to look into the events themselves. So, providing the event name helps show them exactly what the user is doing to abuse the product. For example, they may be creating a new admin for the purpose of privilege escalation.

Explainability also served as a means of model improvement and debugging. When running a trial version, investigators noticed trends in the explained features leading to false positives, prompting changes to feature engineering and risk factor tiers, with some factors lowered, removed, or added based on their prevalence in false positives.

\section{Evaluation and Results}

The model was evaluated using two methods: a counterfactual replay against a historical corpus of known product abuse examples (i.e., testing what the model would have caught had it been live during past incidents), and a live beta release against never-before-seen data. Both evaluations showed significant efficacy gains and reduced alert volume vs. existing rules-based alerts.

\subsection{Evaluation Against Historical Corpus}

Using the labels discussed earlier, a historical labeled corpus of 28 true product abuse labels was constructed. The model impact was evaluated for two different purposes against the training corpus: 
    (1) Improve product abuse coverage compared to existing methods, protecting customers more with high-fidelity detections.
    (2) Reduce the number of leads sent to investigators compared to existing methods, minimizing alert fatigue for investigators.

To evaluate the model's performance across these 2 areas, the chosen metrics were recall and average monthly leads. Recall, also known as True Positive Rate, is the percentage of known product abuse incidents in the historical sample dataset that the model captures ($\text{Recall} = TP / (TP + FN)$).
Average monthly leads is simply the number of leads sent divided by the number of months in the sample dataset. Investigators can only handle so many leads before becoming fatigued; therefore, this was controlled ahead of time with a conservative flag rate for sending leads. Fewer monthly leads along with higher recall represent a successful model in both facets.

Table \ref{tab:results_tab} shows how the model catches more true product abuse events while flagging fewer leads than existing rules-based methods (baseline). The product abuse recall increased 35\% vs the baseline (95\% CI [15, 56] pts, paired Wald interval), and average monthly leads decreased 30\% (95\% CI [22\%, 37\%], exact Poisson rate-ratio method \cite{przyborowski1940homogeneity}), exceeding goals for both efficacy and alert reduction.

\begin{table}[!h]
    \centering
    \begin{tabular}{||ccc||}
    \hline
        Stat & Model & Rules-Based \\
        \hline\hline
        Recall & \textbf{39\% [24, 58]} & 4\% [1, 18]\\
        \hline
        Avg Monthly Lead Reduction & \textbf{30\% [22, 37]} & NA\\
        \hline
    \end{tabular}
    \caption{Using a historical corpus of 28 product abuse cases (3-month window), the new approach catches 9.8$\times$ more instances. Bracketed values are 95\% CIs (Wilson score interval \cite{wilson1927probable} for recall; exact Poisson rate-ratio \cite{przyborowski1940homogeneity} for lead reduction). This is a significant operational advantage, mitigating what was previously a reactive-only issue.
    }
    \label{tab:results_tab}
\end{table}

\subsection{Beta Mode Evaluation}

As an additional evaluation and learning step, the model was released in ``beta mode'', where it predicted against live, never-before-seen data. During the beta release, investigators labeled predictions in a limited capacity vs existing queues. Since this wasn't reliant on a single labeled corpus but on two queues with very different proportions of alerts reviewed (the new model vs the rule-based method), efficacy was measured by precision and alert volume. \textbf{Investigators reviewed 1,925 rules-based alerts vs 173 ML alerts in the time period of ~5 months}, with ML alerts showing a significant precision increase and alert volume decrease vs existing rules alerts. 
Efficacy was measured by precision ($\text{Precision} = TP / (TP + FP)$) and alert volume.

The average monthly leads metric was used again to calculate the average monthly leads. For this evaluation, the beta model ran for 5 months against live data.

Table \ref{tab:results_tab_beta} shows how the beta model, despite having far fewer alerts reviewed, exceeded rules precision by 9.6 pts (95\% CI [4.5, 14.7]) and reduced alerts fired by 57.6\% (95\% CI [54.3\%, 60.7\%]) compared to rules-based alerts. In total, 173 ML alerts were reviewed, resulting in 23 TPs (19 of which were novel, missed by the high/critical rule queue); 1,925 rules-based alerts were reviewed, with 71 TPs discovered.

\begin{table}[!h]
    \centering
    \begin{tabular}{||ccc||}
    \hline
        Stat & Model & Rules-Based \\
        \hline\hline
        Precision & \textbf{13.3\% [9.0, 19.2]} & 3.7\% [2.9, 4.6]\\
        \hline
        Avg Monthly Lead Reduction & \textbf{57.6\% [54.3, 60.7]} & NA\\
        \hline
    \end{tabular}
    \caption{Our ML solution was deployed against live data in a 5-month ``beta'' release, showing far higher precision than the rule-based approach. Bracketed values are 95\% CIs (Wilson score interval \cite{wilson1927probable} for precision; exact Poisson rate-ratio \cite{przyborowski1940homogeneity} for lead reduction). Given the high-risk and low-enough-to-staff load of alerts, the model is continuing expanded roll out due to its success.
    }
    \label{tab:results_tab_beta}
\end{table}

The beta model was also very beneficial for model debugging and improvements. Due to the explainability aspects, trends were identified that tended to produce false positives. Some of these originated from specific risk factors being hit when they shouldn't have been, which led to modifying that field upstream. Similar trends were spotted, and modifications were made to feature engineering and risk factor bucketing. 

The model is also adaptable to new labels, as we collect these new labels over time and can build an updated corpus with them. The risk score weights can be automatically re-tuned based on this labeled corpus, additional risk factors may be added, problematic ones may be dropped, etc. 

Due to the transparency and explainability of each detection, it is often apparent to investigators why it is a true positive or false positive and whether that indicates a one-time failure or a symptom of the model that could be adjusted. Overall, explainability is not only a tool for investigators to understand where to start their investigation, but it is also a tool for filtering out false positives and incorporating feedback into model adjustments down the line. 

\section{Conclusion}

By combining multiple data streams covering different time scales and update frequencies, we build an effective strategy for identifying product abuse that far exceeds rules-based approaches. A mix of unsupervised anomaly detection is combined with domain-expert ordinal coding to devise a final risk score that enables finding product abuse within the data 10$\times$ more effectively. Incorporating explainability enabled rapid iteration and information retrieval by investigators to improve the design. 

\section*{Acknowledgment}
The authors used Claude \cite{claude2026} throughout the preparation of this manuscript to assist with language editing, phrasing, LaTeX formatting, and revisions. All technical content, experimental design, results, and conclusions are the authors' own.

\bibliographystyle{IEEEtran}
\bibliography{references,sample-base}

\clearpage

\appendix

\section{Charts and Additional Data}
\begin{figure}[!h]
    \centering
    \includegraphics[width=1\linewidth]{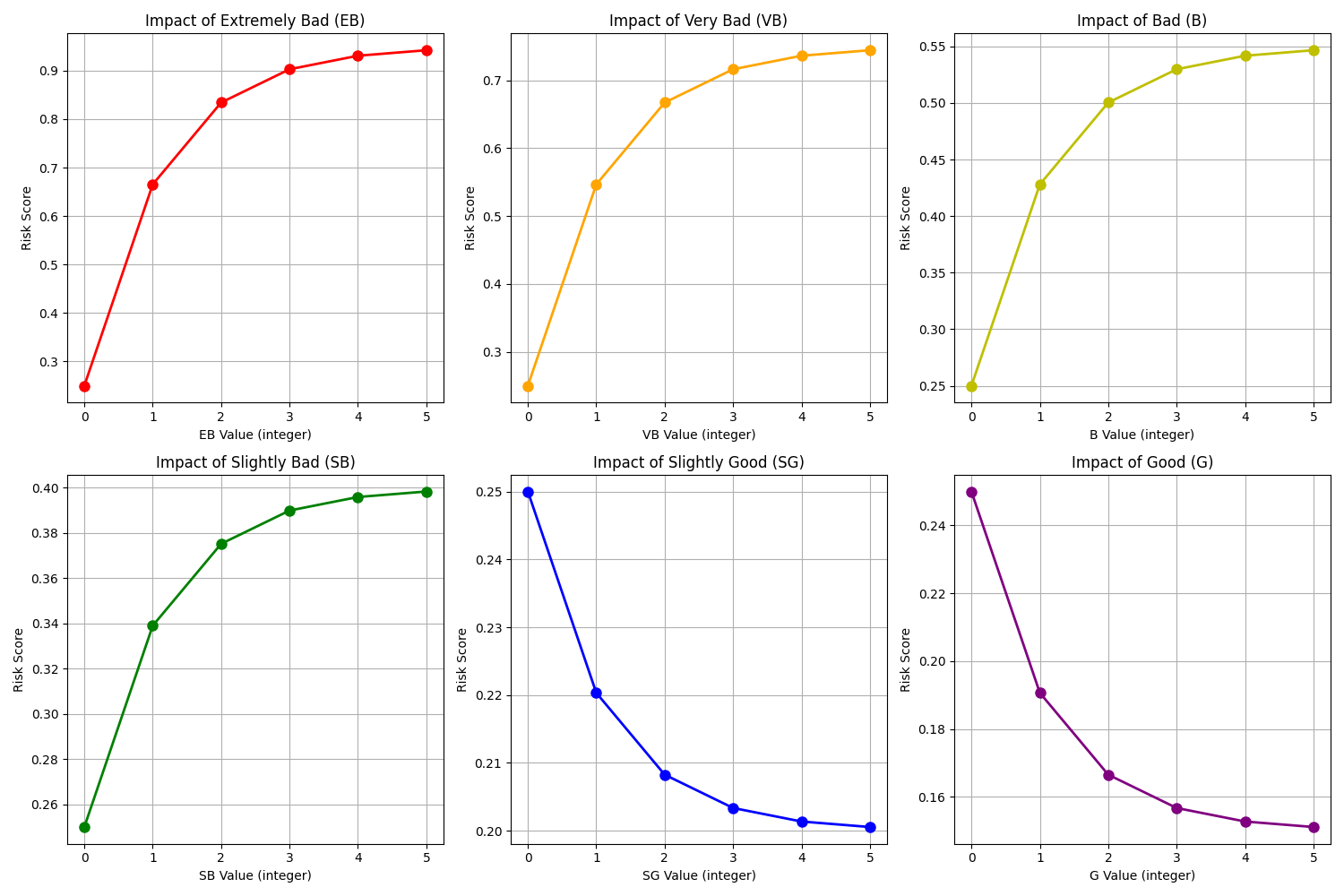}
    \caption{Risk Score Diagram}
    \label{fig:risk_score_diag4}
\end{figure}

\end{document}